\documentclass[twocolumn,showpacs,preprintnumbers,amsmath,amssymb,prl]{revtex4}
\usepackage{graphicx}
\usepackage{bm}
\begin{document}

\title{Magnetotransport in ferromagnetic Mn$_5$Ge$_3$, Mn$_5$Ge$_3$C$_{0.8}$, and Mn$_5$Si$_3$C$_{0.8}$ thin films}
\author{Christoph S\"{u}rgers$^{1}$}
\email{Christoph.Suergers@kit.edu}
\author{Gerda Fischer$^{1}$}
\author{Patrick Winkel$^{1}$}
\author{Hilbert v. L\"ohneysen$^{1,2}$}
\affiliation{$^1$Karlsruhe Institute of Technology, Physikalisches Institut and DFG-Center for Functional Nanostructures, P.O. Box 6980, 76049 Karlsruhe, Germany}
\affiliation{$^1$Karlsruhe Institute of Technology, Institut f\"ur Festk\"orperphysik, P.O. Box 3640, 76021 Karlsruhe, Germany}
\begin{abstract}
The electrical resistivity, anisotropic magnetoresistance (AMR), and anomalous Hall effect of ferromagnetic Mn$_5$Ge$_3$, Mn$_5$Ge$_3$C$_{0.8}$, and Mn$_5$Si$_3$C$_{0.8}$ thin films has been investigated. The data show a behavior characteristic for a ferromagnetic metal, with a linear increasee of the anomalous Hall coefficient with Curie temperature. While for ferromagnetic Mn$_5$Si$_3$C$_{0.8}$ the normal Hall coefficient $R_0$ and the AMR ratio are independent of temperature, these parameters strongly increase with temperature for the germanide films. This difference is attributed to the different hybridization of electronic states in the materials due different lattice parameters and different atomic configurations (Ge vs. Si metalloid). The concomitant sign change of $R_0$ and the AMR ratio with temperature observed for the germanide films is discussed in a two-current model indicating an electron-like minority-spin transport at low temperatures.    
\end{abstract}
\date{\today}

\maketitle

\section{I. Introduction}
The vision of spintronics - the development of faster and less power-consuming nonvolatile electronics with increased integration density by utilizing the electron's spin degree of freedom - strongly depends on the ability to inject, manipulate, and detect spin-polarized charge carriers in the semiconductor  \cite{wolf_spintronics:_2001,zutic_spintronics_2004,jansen_silicon_2012}. In search of new materials for spintronic applications, a number of ferromagnetic metals and compounds are   being explored with the aim to overcome the various obstacles of spin injection and detection in semiconductors, in particular in Si, and in ferromagnet-semiconductor heterostructures.  Ferromagnetic silicides or germanides are favorable due to the possible integration into semiconductor Si- and Ge-based electronics and complementary metal-oxide-semiconductor (CMOS) technology \cite{zhou_electrical_2011}. Mn$_5$Ge$_3$ films are an example because they can be epitaxially grown on Ge(111) and are ferromagnetic at room temperature with a Curie temperature $T_{\rm C}$ = 296 K \cite{zeng_epitaxial_2003} close to $T_{\rm C}$ = 304 K of bulk Mn$_5$Ge$_3$ \cite{kanematsu_convalent_1962}. For these films, the intrinsic part of the extraordinary or anomalous Hall effect (AHE) has been shown to depend linearly on the magnetization \textit{M} \cite{zeng_linear_2006}. However, for a real device operating at room temperature $T_{\rm C}$ values well above room temperature are pivotal. This can be achieved, e.g., by inserting carbon atoms into Mn$_5$Ge$_3$ \cite{gajdzik_strongly_2000,slipukhina_simulation_2009,spiesser_control_2011}. Recently, Mn$_5$Ge$_3$C$_{0.8}$ has been implemented in MOS capacitors and Schottky diodes on n-Ge to determine work functions and contact resistivities \cite{fischer_ferromagnetic_2013}. Furthermore, Mn$_5$Ge$_3$/Ge and Mn$_5$Ge$_3$C$_{0.8}$/Ge heterostructures are being investigated for potential spintronic applications \cite{thanh_epitaxial_2013}.

A stabilization of ferromagnetic order by carbon has also been established for the prototype material Mn$_5$Si$_3$ which orders antiferromagnetically below 100 K but can be driven ferromagnetic by insertion of carbon with $T_{\rm C} \approx$\, 350 K for Mn$_5$Si$_3$C$_{0.8}$ \cite{surgers_preparation_2003,gopalakrishnan_electronic_2008,surgers_magnetism_2009}. The high $T_{\rm C}$ well above room temperature makes this material interesting to study in light of potential applications in combination with silicon, the mainstream semiconductor. A previous electronic-transport study performed on Mn$_5$Si$_3$C$_{x}$ focused on the effect of carbon concentration $x$ and film thickness $d$ on the resistivity, where the carbon-induced disorder gives rise to scattering of electrons by structure-induced two-level systems at low temperatures \cite{gopalakrishnan_electronic_2008}.   

In ferromagnetic materials, the spin-orbit interaction (SOI) gives rise to an anisotropic magnetoresistance (AMR). The AMR is the difference between the magnetoresistance (MR) when the magnetization $M$ is aligned in the longitudinal (L) or transverse (T) direction with respect to the current, and the field is oriented in the plane of the film. In 3\textit{d} transition metals the AMR ratio $\Delta \rho / \rho = (\rho_{\parallel {\rm ,L}}-\rho_{\parallel {\rm ,T}}) /\rho_{\parallel {\rm ,T}}$ is usually a few percent and often larger than the ordinary MR which is caused by the Lorentz force acting on the charge carriers and also observed in nonmagnetic metals. Both effects are linked by the microscopic electronic properties of the material such as the spin-split band structure, the density of states (DOS), and the SOI.

AMR and AHE have been known for almost a century \cite{nagaosa_anomalous_2010,mcguire_anisotropic_1975,campbell_transport_1982,fang_anomalous_2003} and experienced a renaissance in recent years. Separating the ordinary and anomalous Hall coefficients in ferromagnetic films requires measuring Hall voltage, MR, and $M$ simultaneously \cite{nagaosa_anomalous_2010}. 

Although the structural and magnetic properties of ferromagnetic Mn$_5$Ge$_3$C$_x$ and Mn$_5$Ge$_3$C$_x$ films have been investigated previously, a detailed magnetotransport study of  these films is lacking. Hence, we have conducted a comprehensive investigation of the magnetotransport properties of ferromagnetic Mn$_5$Ge$_3$, Mn$_5$Ge$_3$C$_{0.8}$, and Mn$_5$Si$_3$C$_{0.8}$ films for temperatures 2 - 400 K. In the germanide films we find a strong temperature dependence of the AMR ratio and of the ordinary Hall coefficient $R_0$ which both change sign from negative to positive with increasing temperature. In contrast, temperature independent positive $R_0$ and AMR ratio are observed for ferromagnetic Mn$_5$Si$_3$C$_{0.8}$. We argue that the difference between the germanide and silicide films arises from the variation of the spin-split band structure and hybridization in these materials due to the different lattice parameters that sensitively affect the electronic and magnetic properties.

\subsection{A. Materials properties}
The prototype phase of the investigated films is the intermetallic compound Mn$_5$Si$_3$ with D8$_8$ structure. The hexagonal unit cell (space group $P{\rm 6}_3/mcm$) contains two formula units with 10 Mn atoms on two inequivalent lattice sites (4 Mn$_1$, 6 Mn$_2$) and 6 Si atoms. The antiferromagnetic structure of Mn$_5$Si$_3$ has been determined by neutron diffraction \cite{brown_low-temperature_1992,brown_antiferromagnetism_1995,gottschilch_study_2012} uncovering a non-collinear spin structure below 68 K which gives rise to a topological Hall effect \cite{surgers_large_2014}. Inserting carbon atoms to yield Mn$_5$Si$_3$C$_x$, gives rise to an anisotropic modification of the local structure around the Mn sites and induces ferromagnetic order with a maximum $T_{\rm C}$ = 352 K for $x$ = 0.8 \cite{gajdzik_ferromagnetism_2000,surgers_preparation_2003}. Site-dependent magnetic moments averaging to  1.19 $\mu_{\rm B}/{\rm Mn}$ have been inferred for ferromagnetic Mn$_5$Si$_3$C from \textit{ab-initio} calculations and a local moment of 1.9 $\mu_{\rm B}$ attributed to Mn$_2$ has been observed by broad-band nuclear magnetic resonance \cite{surgers_local_2002}.  

The isostructural Mn$_5$Ge$_3$ compound is ferromagnetic with a Curie temperature $T_{\rm C} =$ 304 K \cite{kanematsu_convalent_1962,forsyth_spatial_1990}. \textit{Ab-initio} calculations indicate the presence of two competing magnetic phases, a collinear phase and a phase with small non-collinearity \cite{stroppa_competing_2007}. The ferromagnetic stability can be enhanced by carbon insertion \cite{gajdzik_strongly_2000,surgers_magnetic_2008,spiesser_control_2011} possibly due to a 90$^{\circ}$ ferromagnetic superexchange mediated by C \cite{slipukhina_simulation_2009}. A substantial modification of the electronic band structure due to carbon was also derived from a comparison of the $T_{\rm C}$ dependence on the unit-cell volume for Mn$_5$Si$_3$C$_x$ and Mn$_5$Ge$_3$C$_x$ \cite{surgers_magnetism_2009}. In polycrystalline films, a maximum $T_C \approx$\,450 K was reached for Mn$_5$Ge$_3$C$_{0.8}$ \cite{gajdzik_strongly_2000,surgers_magnetic_2008}. For higher $x$, $T_{\rm C}$ and the magnetization decrease due to the formation of additional phases. A similar C-induced effect was observed for epitaxially grown Mn$_5$Ge$_3$C$_x$ films on Ge (111) substrates with $T_{\rm C}$ = 430 - 450 K for $x \approx 0.7 - 0.8$, making the material an interesting candidate for potential spintronic applications \cite{spiesser_control_2011,thanh_epitaxial_2013}.

\subsection{B. Anomalous Hall effect}
The electrical resistivity $\rho = V wd/LI$ of a film of thickness $d$ and width $w$ is determined from the longitudinal voltage $V$ measured along a stripe of length $L$ with current $I$. The Hall effect is measured as transverse voltage $V_{\rm xy}$ to the current $I$ in perpendicular magnetic field $H$, where the Hall resistivity is obtained via $\rho_{\rm xy} = V_{\rm xy} d / I$. In ferromagnetic materials, the Hall effect comprises the ordinary term $\rho_{\rm xy}^0$ arising from the Lorentz force acting on the charge carriers, and the extraordinary or anomalous term $\rho_{\rm xy}^{AH}$ due to the magnetization $M$ \cite{nagaosa_anomalous_2010}:
\begin{equation}
\rho_{\rm xy} = R_0B + R_S\mu_0M = \rho_{\rm xy}^0 + \rho_{\rm xy}^{AH}.
\label{eqn1}
\end{equation}
$R_0 = (e n_{\rm eff})^{-1}$ ($n_{\rm eff}$: effective carrier density) and $R_S$ are the ordinary and anomalous Hall coefficients, respectively, $\mu_0$ the magnetic constant, $M$ the magnetization and $B = \mu_0 \left[H+M(1-N)\right]$, where the demagnetization factor $N$ for thin films in a perpendicular magnetic field is $N \approx 1$. Hence, $B = \mu_0H$ and the Hall resistivity and Hall conductivity can be written as 
\begin{equation}
\rho_{\rm xy}(H,T) = R_0 \mu_0H + S_H \rho^2(H,T) M(H,T) 
\label{eqn2}
\end{equation}    
\begin{equation}
\sigma_{\rm xy} = \rho_{\rm xy}/\rho^2 = R_0 \mu_0H/\rho^2 + S_H M = \sigma_{\rm xy}^0 + \sigma_{\rm xy}^{AH}
\label{eqn3}
\end{equation}    
where $S_H = \mu_0R_S/\rho^2$ and we have assumed $\rho_{\rm xy} \ll \rho$. The above expressions are valid in weak magnetic fields for which $\omega_{\rm c} \tau \ll 1$, where $\omega_{\rm c} = eB/m$ is the cyclotron frequency, $\tau = m/ne^2\rho$ is the electron scattering time, and $m$ is the effective electron mass. 
 
For a particular system, $R_0$ may change with temperature due to the different contributions from several electron-like and hole-like bands crossing the Fermi surface. The anomalous contribution $\sigma_{\rm xy}^{AH}$ contains an intrinsic contribution originating from the Berry-phase curvature correction to the group velocity of a Bloch electron induced by SOI as well as extrinsic contributions arising from a side-jump mechanism and skew scattering \cite{nagaosa_anomalous_2010}. The intrinsic contribution dominates the AHE in moderately conducting materials while the skew scattering contribution is important at low temperatures and in clean samples of low impurity concentration. A scaling relation $\sigma_{\rm xy} \propto \rho^{-\alpha}$ ($\alpha \ge 0$) has been proposed to cover the different transport regimes \cite{onoda_quantum_2008}. The conventional theories of the AHE derived via pertubation theory have shown $S_H \propto \lambda_{SO}$ independent of $T$, at least below the Curie temperature $T_{\rm C}$ \cite{nagaosa_anomalous_2010,fang_anomalous_2003}. The contributions from $\sigma_{\rm xy}^0$ and $\sigma_{\rm xy}^{AH}$ can be disentangled by measuring the whole set of resistivities $\rho_{\rm xy}(H,T)$ and $\rho(H,T)$ and the magnetization $M(H,T)$. 

\section{II. Experimental}
Thin polycrystalline Mn$_5$Ge$_3$, Mn$_5$Ge$_3$C$_{0.8}$, and Mn$_5$Si$_3$C$_{0.8}$ films were prepared by magnetron sputtering in high vacuum (base pressure $p < 10^{-4}$\, Pa) from elemental targets at substrate temperatures $T_{\rm S}$ = 400 - 470 $^{\circ}$C and were characterized by x-ray diffraction to confirm formation of the Mn$_5$Si$_3$-type structure as described earlier \cite{surgers_preparation_2003,gopalakrishnan_electronic_2008}. The films have a coarsely grained morphology with an average grain size equal to the film thickness. (11\=20) oriented sapphire substrates covered by a mechanical mask were used to obtain a Hall-bar layout. For the samples investigated here, $w$= 0.5 mm, $d$ = 50 nm (Mn$_5$Ge$_3$C$_x$) and 45 nm (Mn$_5$Si$_3$C$_{0.8}$), $L$ = 8 mm. Contacts to the sample were made by attaching thin Cu wires to the film, glued with silver epoxy. Resistivities were measured in a physical property measuring system (PPMS, Quantum Design) for magnetic fields $\mu_0H$ up to $\pm$ 8 T and temperatures 2 - 400 K. The magnetic field was oriented either perpendicularly to the film plane or either longitudinally ($\rho_{\parallel {\rm ,L}}$) or transverse ($\rho_{\parallel {\rm ,T}}$) to the current direction in the film plane. The Hall resistivity $\rho_{\rm xy}$ was obtained by performing a field sweep from negative to positive values, $\rho_{\rm xy} = \left[\rho_{\rm xy}(+H)- \rho_{\rm xy}(-H)\right]/2$. 
\begin{figure}
\centerline{\includegraphics[width=0.9\columnwidth]{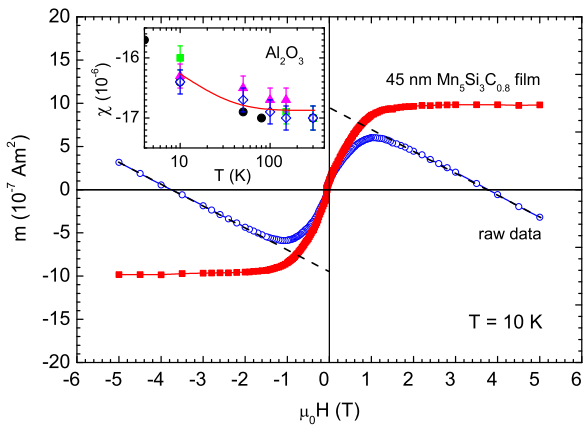}}
\caption[]{Magnetic moment $m$ of a 45-nm Mn$_5$Si$_3$C$_{0.8}$ film (volume $V_f = 1.65\times 10^{-6}\, \rm{cm}^3$) on a (11\=20)-oriented Al$_2$O$_3$ substrate (volume $V_s = 2\times 10^{-2}\, \rm{cm}^3$) at $T$\, = 10 K in perpendicular magnetic field. Dashed lines indicate a linear $m(H)$ behavior. Open symbols indicate raw data, closed symbols indicate the magnetic moment of the ferromagnetic film after the subtraction of the diamagnetic contribution arising from the substrate. Inset: Semilogarithmic plot of the magnetic susceptibility $\chi(T)$ of sapphire, see text for details.}
\label{fig1}
\end{figure}

The magnetic moment $m$ of the films was measured in a superconducting quantum-interference device (SQUID) magnetometer between 10 and 350 K for magnetic fields up to 5 T. For the determination of the sample magnetization, in particular for temperatures close to $T_{\rm C}$ where $M(T)$ does not saturate in magnetic field, a correct subtraction of the diamagnetic signal of the Al$_2$O$_3$ substrate (volume $V_s$) is crucial. As an example, Fig. \ref{fig1} shows raw $m(H)$ data of a film on a Al$_2$O$_3$ substrate measured at 10 K and $m(H)$ of the Mn$_5$Si$_3$C$_{0.8}$  film after substraction of the diamagnetic contribution from the Al$_2$O$_3$\, substrate. For the subtraction we have used the magnetic susceptibility $\chi(T)$ of the substrate indicated by the red line in the inset of Fig. \ref{fig1}. $\chi(T)$ was found to vary between -16.27 ($T$ = 10 K) and -16.87 ($T \ge$ 150 K). The red line is the average of various values $\chi = (\Delta m/\Delta H)/V_s$, determined from the slope $\Delta m/\Delta H$ of the linear $m(H)$ behavior above the saturation field at $T$ = 10 K, for ferromagnetic films of Fe (squares), Mn$_5$Si$_3$C$_{0.8}$ (triangles), and Mn$_5$Ge$_3$C$_{0.8}$ (diamonds) on sapphire substrates. Solid circles represent data of Al$_2$O$_3$ reported by Smith et al. \cite{smith_magnetic_1970}. 

\section{III. Results}

\begin{figure}[h!]
\centerline{\includegraphics[width=0.8\columnwidth]{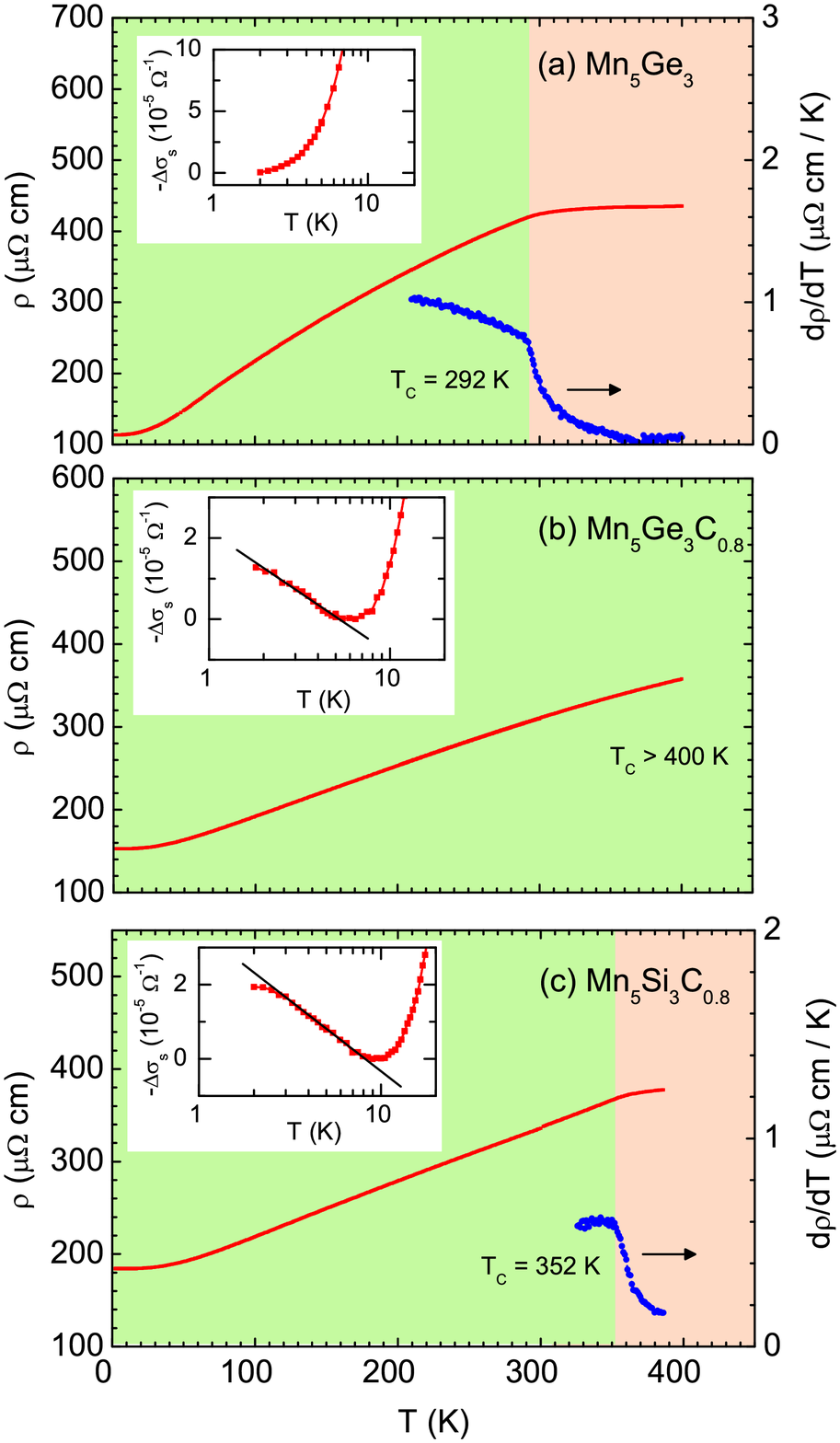}}
\caption[]{Temperature dependence of resistivity $\rho$ (solid line) for ferromagnetic (a) Mn$_5$Ge$_3$, (b) Mn$_5$Ge$_3$C$_{0.8}$, and (c) Mn$_5$Si$_3$C$_{0.8}$ films. Kinks in $d\rho(T)/dT$ in the vicinity of the Curie temperature $T_{\rm C}$ are shown in (a) and (c). Upper insets show a semilogarithmic plot of the variation of the sheet conductance $-\Delta\sigma_s(T) = [\rho(T)-\rho(T_{0})]/\rho^2(T_{0})$, where the solid line indicates a behavior $-\Delta \sigma_s(T) \propto {\rm log}(T/T_{0}$).}
\label{fig2}
\end{figure}

The temperature dependence of the resistivity $\rho$ for the three films is shown in Fig. \ref{fig2}. At the lowest temperature, the films have residual resistivities in the range 100 - 200 $\mu \Omega$cm and exhibit at higher temperatures a roughly linear temperature dependence characteristic for a metal. 
\begin{figure*}
\centerline{\includegraphics[width=2\columnwidth]{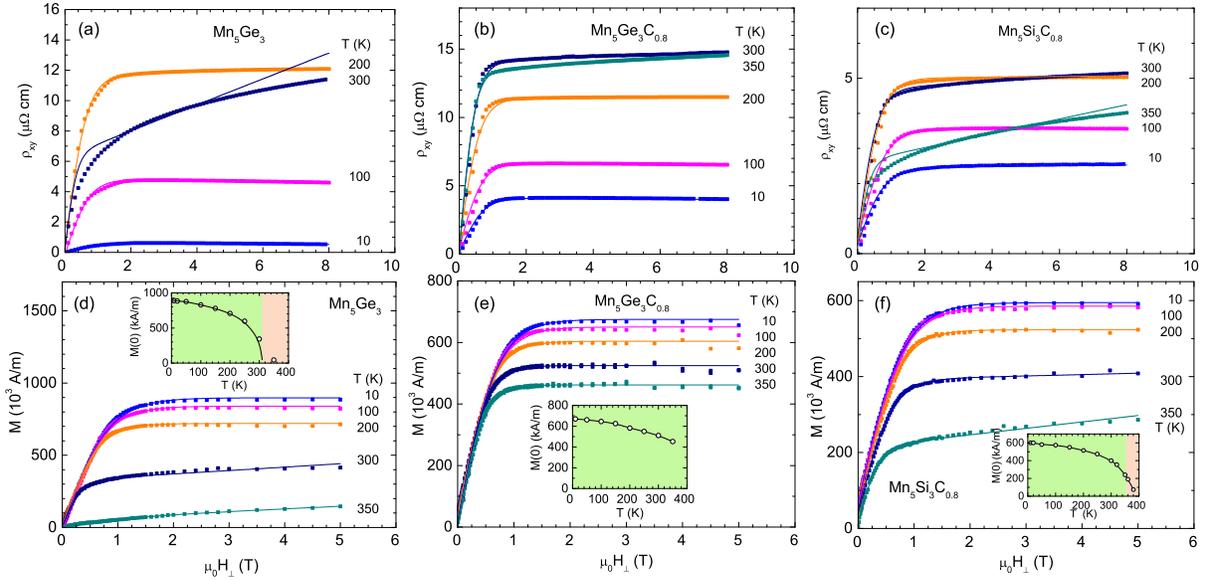}}
\caption[]{(a-c) Hall resistivity $\rho_{\rm xy}$, and (d-f) magnetization $M$ in perpendicular field $H_{\perp}$ at various temperatures $T$. Solid lines show fits according to Eqn. \ref{eqn2} to measured data (symbols), see text for details. Insets show the temperature dependence of the magnetization $M(0)$ obtained from the linear extrapolation of the high-field magnetization $M(H_{\perp})$ toward $H_{\perp} \rightarrow 0$.} 
\label{fig3}
\end{figure*}
The temperature dependence is in agreement with previously published data of Mn$_5$Si$_3$C$_{0.8}$ films including a logarithmic $T$ dependence of the sheet conductance (insets) \cite{gopalakrishnan_electronic_2008}. 

For the C-inserted films this behavior is attributed to the scattering of conduction electrons by two-level systems originating from C-induced disorder, with a crossover to Fermi-liquid behavior below $\approx$ 1 K. At high temperatures, the slope of $\rho(T)$ changes and a weak kink appears at the Curie temperature $T_{\rm C}$ indicated in (a) and (c). This is not observed for Mn$_5$Ge$_3$C$_{0.8}$ consistent with a $T_{\rm C} \approx 450$ K of this compound \cite{gajdzik_strongly_2000,spiesser_control_2011}, which was not accessible by the experimental set-up used in this study. The magnetic phase transitions are better resolved in the derivative $d\rho(T)/dT$, see Fig. \ref{fig2} (a,c), which shows ca clear jump at $T_{\rm C}$  \cite{zumsteg_electrical_1970}. The $T_{\rm C}$ values determined from $d\rho(T)/dT$ are in very good agreement with earlier published data for Mn$_5$Ge$_3$ \cite{zeng_epitaxial_2003}, Mn$_5$Ge$_3$C$_{0.8}$ \cite{gajdzik_strongly_2000,spiesser_control_2011,thanh_epitaxial_2013}, and Mn$_5$Si$_3$C$_{0.8}$ films \cite{surgers_preparation_2003,gopalakrishnan_electronic_2008}.

Due to the high residual resistivities and low resistance ratios $RRR \approx 2-4$ the films fall into the intrinsic Hall-effect regime \cite{onoda_quantum_2008}. From $\rho_0 l = 4.25 \times 10^{-15}\, \Omega{\rm m}^2$ for Mn$_5$Ge$_3$ \cite{panguluri_spin_2005} an electron mean free path $l \approx$\, 3 nm is estimated, much smaller than the film thickness. Therefore, finite-size effects arising from electron scattering at the film boundaries are considered to be negligible.  

Figure \ref{fig3} (a-c) show the Hall resistivity $\rho_{\rm xy}(H)$ of the compounds for different $T$. For clarity, only a subset of data is shown. $\rho_{\rm xy}(H)$ of the ferromagnetic films shows a steep increase with field at low fields and a saturation at high fields for $T \ll T_{\rm C}$, resembling the magnetization behavior $M(H)$, see Fig. \ref{fig3}(d-f). We do not observe a non-linear behavior of $\rho_{\rm xy}(H)$ or a sign change with magnetic field that would allow a separation of electron and hole contributions \cite{watts_evidence_2000}. This indicates that field-induced changes of particular orbits or a reconstruction of the Fermi surface are negligible. In perpendicular magnetic field, the magnetization exhibits a hard-axis behavior without hysteresis due to the strong shape anisotropy of the thin film. The magnetization $M(0)$ determined by extrapolating the high-field $M(H)$ behavior to $H = 0$ shows the characteristic dependence of a ferromagnet (see insets). $M(0)$ is zero at $T_{\rm C}$ obtained from the jump in $d\rho(T)/dT$. For the saturation magnetization we obtain $M_S\rm{(10\, K)} = 6\times 10^5 {\rm A/m}$ corresponding to a magnetic moment of $1.3\, \mu_B/{\rm Mn}$, somewhat higher than observed earlier for 100-nm Mn$_5$Si$_3$C$_{0.8}$ films but similar to 400-nm thick C-implanted films \cite{gopalakrishnan_electronic_2008,surgers_magnetic_2008}. For Mn$_5$Ge$_3$C$_{0.8}$, $M_S\rm{(10\, K)} = 6.7 \times 10^5 {\rm A/m}$ ($1.6\, \mu_B/{\rm Mn}$), 27 \% lower than for 400-nm thick implanted films ($2.2.\, \mu_B/{\rm Mn}$) \cite{surgers_magnetic_2008}. For the Mn$_5$Ge$_3$ film we obtain $M_S\rm{(10\, K)} = 9 \times 10^5 {\rm A/m}$ ($2.1\, \mu_B/{\rm Mn}$), 20 \% lower than $M_S$ of bulk Mn$_5$Ge$_3$ ($2.6\, \mu_B/{\rm Mn}$). Apart from the fact that the 50-nm films have a low magnetic moment which is difficult to measure, the reduced magnetization compared to thick films or bulk is presumably due to a magnetically disordered layer, possibly close to the substrate/film interface.
\begin{figure*}
\centerline{\includegraphics[width=2\columnwidth]{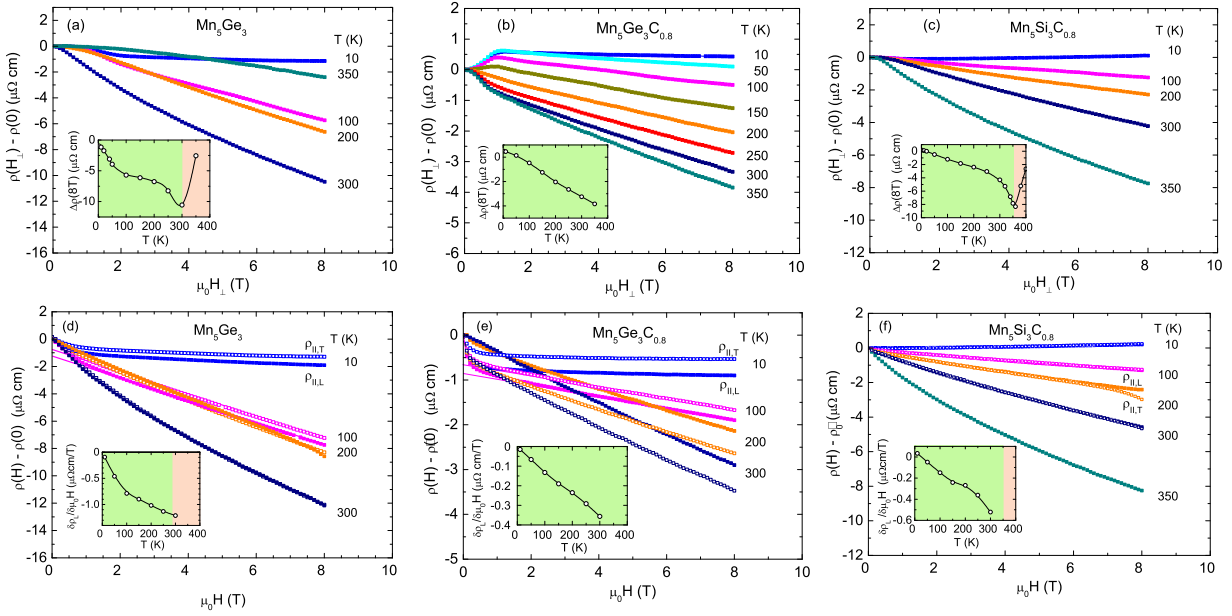}}
\caption[]{(a-c) Magnetoresistivity $\rho(H_{\perp})$ in perpendicular magnetic field for various temperatures $T$. Insets show the temperature dependence of $\Delta \rho = \rho(\rm{H_{\perp}= 8\, T})-\rho(0)$. (d-f) $\rho(H)$ with the magnetic field oriented in the plane and either longitudinal ($\rho_{\parallel {\rm ,L}}$, closed symbols) or transverse ($\rho_{\parallel {\rm ,T}}$, open symbols) to the direction of the current. Insets show temperature dependence of the nearly linear slope $\delta \rho_{\parallel {\rm ,L}}/(\delta \mu_0H$) of the longitudinal MR. Solid lines indicate the extrapolation of the MR to zero field for the determination of the AMR ratio.}
\label{fig4}
\end{figure*}

\begin{figure}
\centerline{\includegraphics[width=0.8\columnwidth]{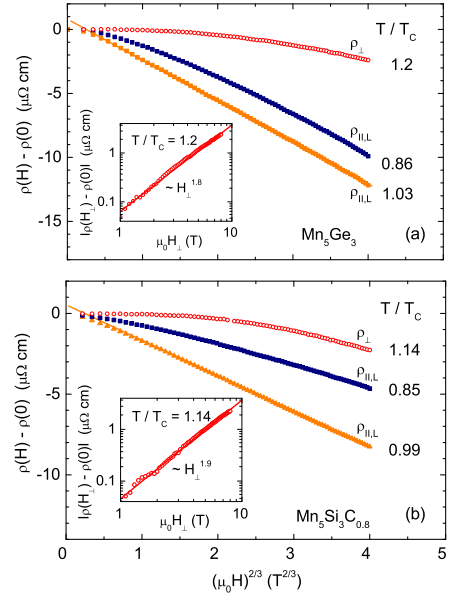}}
\caption[]{MR of (a) Mn$_5$Ge$_3$ ($T_{\rm C}$ = 292 K) and (b) Mn$_5$Si$_3$C$_{0.8}$ ($T_{\rm C}$ = 352 K) vs. $H^{2/3}$ for temperatures close to $T_{\rm C}$. $T_{\rm C}$ could not be reached for Mn$_5$Ge$_3$C$_{0.8}$. $\rho_{\perp}$ indicates data measured in perpendicular field, $\rho_{\parallel {\rm ,L}}$ the longitudinal MR with the field in the plane. Insets show double-logarithmic plots of the MR vs. perpendicular magnetic field $H_{\perp}$ just above $T_{\rm C}$. Solid lines indicate a power-law behavior.}  
\label{fig5}
\end{figure}

The magnetoresistance (MR) is negative for all temperatures as shown in Fig. \ref{fig4}, except for Mn$_5$Ge$_3$C$_{0.8}$ where a small positive MR is observed in a weak perpendicular magnetic field at temperatures $T \le$\,150 K. Changes of the MR at low fields $\mu_0H <$\,1 T are attributed to a change of the magnetic domain structure. In perpendicular field the relative change $\Delta\rho = \rho(8\, \rm{T})-\rho(0)$ varies with temperature, see insets Fig. \ref{fig4}(a-c). $\Delta\rho$ decreases with increasing temperature all the way up to $T_{\rm C}$ and increases again with a distinct minimum at $T_{\rm C}$ observed in Fig. \ref{fig4}(a) and (c). The negative MR in perpendicular field was reported earlier for Mn$_5$Si$_3$C$_{0.8}$ \cite{gopalakrishnan_electronic_2008} and was attributed to the damping of spin-waves by the magnetic field \cite{raquet_electron-magnon_2002}. In a high magnetic field a gap opens in the magnon spectrum and the electron-magnon scattering is suppressed leading to a decrease of the resistivity. Close to $T_{\rm C}$, the MR shows a non-linear behavior, MR $\propto H^{2/3}$ for $T < T_{\rm C}$ and MR $\propto H^{\alpha}$ with $\alpha$\,= 1.8 - 1.9 for $T > T_{\rm C}$, for Mn$_5$Si$_3$C$_{0.8}$ and Mn$_5$Ge$_3$ as shown in Fig. \ref{fig5} where the MR is plotted vs. $H^{2/3}$. This is in qualitative agreement with a simple model where a localized spin system is approximated by a molecular field and the MR is due to $s-d$ scattering \cite{yamada_magnetoresistance_1973}.        

\begin{figure}
\centerline{\includegraphics[width=1.1\columnwidth]{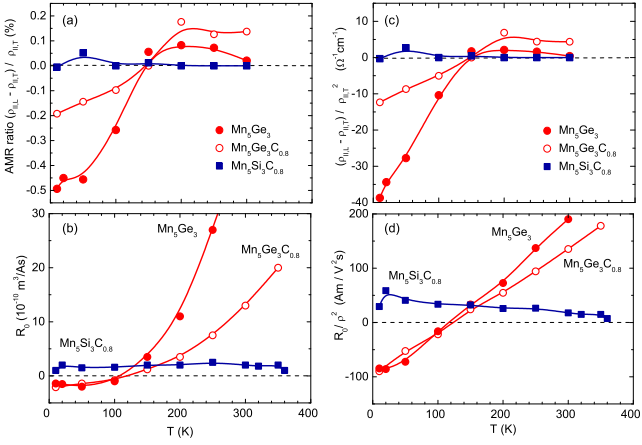}}
\caption[]{(a,b) Temperature dependence of the AMR ratio and of the ordinary Hall coefficient $R_0$, respectively, for Mn$_5$Ge$_3$ (closed circles), Mn$_5$Ge$_3$C$_{0.8}$ (open circles), and Mn$_5$Si$_3$C$_{0.8}$ (squares). (c) and (d) show the reduced AMR ratio $(\rho_{\parallel {\rm ,L}}-\rho_{\parallel {\rm ,T}})/\rho_{\parallel {\rm ,T}}^2$ and the reduced Hall coefficient $R_0/\rho^2$, respectively.}
\label{fig7}
\end{figure}

\begin{figure}
\centerline{\includegraphics[width=0.8\columnwidth]{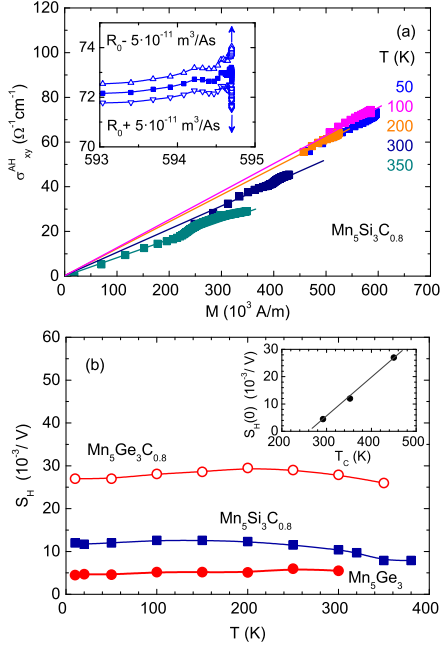}}
\caption[]{(a) Anomalous contribution $\sigma^{\rm AH}_{\rm xy}$ vs. $M$ for Mn$_5$Si$_3$C$_{0.8}$. Colors indicate different temperatures, cf. Figs. \ref{fig3}, \ref{fig4}. Solid lines indicate a linear behavior $\sigma^{\rm AH}_{\rm xy} \propto M$. Inset shows the effect of a variation of the Hall coefficient $R_0$ on the $\sigma^{\rm AH}_{\rm xy}(M)$ behavior at large $M$. (b) Temperature dependence of the anomalous Hall coefficient $S_H$. Inset shows the dependence of $S_H(T \rightarrow 0)$ from the Curie temperature $T_{\rm C}$. Solid line indicates a linear behavior.}
\label{fig6}
\end{figure}

We observe only slight differences between the longitudinal and transverse MR with the field oriented in the plane of the film, see Fig. \ref{fig4}(d-f). Orbital contributions to $\rho_{\parallel {\rm ,L}} \propto (\omega_c \tau)^2 \propto \mu H^2$ are negligibly small ($\mu$: mobility, see below) \cite{porter_scattering_2014}. Similar to the behavior of $\Delta\rho(H_{\perp}$ the slope $\delta\rho / \delta H$ continuously decreases with increasing $T$, see insets. From $\rho_{\parallel {\rm ,L}}$ and $\rho_{\parallel {\rm ,T}}$ we determine the AMR ratio $(\rho_{\parallel {\rm ,L}}-\rho_{\parallel {\rm ,T}})/\rho_{\parallel {\rm ,T}}$ plotted in Fig. \ref{fig7}(a). While a very small AMR almost independent of temperature is observed for Mn$_5$Si$_3$C$_{0.8}$, the AMR of Mn$_5$Ge$_3$ and Mn$_5$Ge$_3$C$_{0.8}$ strongly dependends on temperature. The negative AMR at low temperature increases with increasing temperature up to positive values at high temperatures thereby crossing zero around 150 - 200 K. 

With the data of Figs. \ref{fig3} and \ref{fig4} we are able to separate the different contributions to the Hall effect. We apply Eqns. \ref{eqn2}, \ref{eqn3} to analyze the AHE of the ferromagnetic films. Due to the high residual resistivity, the cyclotron resonance frequency is $\omega_c\tau = R_0 B/\rho \approx 10^{-4} B{\rm (T)} \ll 1$ and the weak-field expressions (no closed cycles) Eqns. \ref{eqn2}, \ref{eqn3} are applicable. 

In Fig. \ref{fig6}(a), $\sigma_{\rm xy}^{AH} = (\rho_{\rm xy} - R_0 \mu_0H)/\rho^2$ at different $T$ is plotted vs. $M$ for Mn$_5$Ge$_3$C$_{0.8}$ as an example, cf. Eqn. \ref{eqn3}. For clarity, again only a subset of data is shown. $R_0$ was used as a free parameter to yield a linear dependence $\sigma_{\rm xy}^{AH}(M) =S_HM$ crossing the origin \cite{jiang_scaling_2010}. This assumption is derived from the linear dependence $\sigma_{\rm xy}^{AH} \propto M$ reported earlier for epitaxial Mn$_5$Ge$_3$ films on Ge(111) and  attributed to the existence of long-wavelength spin fluctuations in this material \cite{zeng_linear_2006}. $R_0$ can be determined with sufficient accuracy because small variations of $R_0$ drastically change the $\sigma_{\rm xy}^{AH}(M)$ behavior, in particular above the saturation field, see inset Fig. \ref{fig6}(a). The influence of the MR on the Hall effect cancels by this procedure. We obtain the Hall coefficients $R_0$ and $S_H$ from the slope of $\sigma_{\rm xy}^{AH}(M)$ allowing calculation of the Hall resistivity $\rho_{\rm xy}(H)$ (Eqn. \ref{eqn2}) for comparison with the experimental data. We obtain good agreement between the measured Hall resistivity and the calculated values, see Fig. \ref{fig3} (a-c), except for temperatures  close to $T_{\rm C}$. We mention that similar values for $R_0$ and $S_H$ are obtained from a plot $\rho_{\rm xy}/\mu_0H$ vs. $\rho^2 M/\mu_0H$. Moreover, adding a contribution $ \propto \rho$ due to skew scattering to the Hall effect does not improve the agreement between the measured and calculated values. This is due to the fact that the resistivities of the polycrystalline films are high and the Hall effect is dominated by the contributions $\propto \rho^2$ \cite{onoda_quantum_2008,nagaosa_anomalous_2010}. 

The coefficients $S_H$ of the anomalous Hall effect determined by this method are plotted in Fig. \ref{fig6}(b). For all three films, the coefficient $S_H$ is positive and $T$-independent almost up to $T_{\rm C}$ due to $S_H \propto \lambda_{SO}$ as observed earlier  \cite{fang_anomalous_2003,nagaosa_anomalous_2010,zeng_linear_2006}. $S_H$ only gradually decreases close to $T_{\rm C}$ but a finite $S_H$ is still observed in the paramagnetic regime above $T_{\rm C}$ presumably due to the $T$-independent spin-orbit interaction $\lambda_{\rm SO}$ \cite{fang_anomalous_2003,tatara_chirality-driven_2002}. $\lambda_{\rm SO}$ can be roughly estimated from the dimensionless coupling for $d$ orbitals of size $r_d \approx$\, 0.05 nm ($Ze^2/2\epsilon_0mc^2r_d$) and the band kinetic energy ($\hbar^2/2ma^2$) \cite{ye_berry_1999}. For $Z_{\rm Mn}$ = 25, $a$ = 0.5 nm we obtain $\lambda_{\rm SO} \approx 0.1\, $meV (1.2 K). $S_H$ successively increases from Mn$_5$Ge$_3$, Mn$_5$Si$_3$C$_{0.8}$, to Mn$_5$Ge$_3$C$_{0.8}$, possibly due to the increasing ferromagnetic stability. This is supported by a linear increase of $S_H (T\rightarrow 0)$ with $T_{\rm C}$ of the samples shown in the inset of Fig. \ref{fig6}(b).

For Mn$_5$Si$_3$C$_{0.8}$, the ordinary Hall coefficient $R_0 \approx 2\times 10^{10} {\rm m}^3/{\rm As}$ is positive and independent of temperature, and corresponds to $n_{\rm eff} = 3\times 10^{22}\, {\rm holes/cm}^{3}$, i.e., a factor five higher than for Mn$_5$Si$_3$, suggesting \textit{p}-type doping by carbon, see Fig. \ref{fig7}(c). This can be due to a carbon-induced change of the electronic band structure and an increased density of states at the Fermi level, similar to what has been found for Mn$_5$Ge$_3$C$_{x}$ \cite{slipukhina_simulation_2009}. In contrast, the ordinary Hall coefficient $R_0$ for the germanide films strongly varies with temperature. In particular, $R_0$ is negative at low $T$ with $R_0(10\, {\rm K}) =-2 \times 10^{-10}\, {\rm m}^3/{\rm As}$ for both Mn$_5$Ge$_3$C$_{x}$ films yielding an effective charge carrier density $n_{\rm eff} = \left|1/R_0 e\right| = 3\times 10^{22}\,{\rm cm}^{-3}$ corresponding to $\approx 0.8$ electrons per Mn and a Hall mobility $\mu = \left|R_0\right|/\rho = 2\, {\rm cm}^2/{\rm Vs}$. This low Hall mobility confirms our statement above that the orbital contribution to $\rho_{\parallel {\rm ,L}}$ is small \cite{porter_scattering_2014}. $R_0$ increases $\propto T^2$ and changes sign around 120 K indicating an increasing contribution from hole-like bands. We note that $R_0$ vs. $T/T_{\rm C}$ obeys a similar behavior for both germanide samples. The temperature dependence of $R_0$ is in agreement with the behavior of epitaxially grown Mn$_5$Ge$_3$ films, where $R_0 \approx -3\times10^{-10} {\rm m}^3/{\rm As}$ was obtained at low temperature with a sign change from negative to positive at 180 K \cite{zeng_linear_2006}. A sign change of $R_0$ was also reported for nonmagnetic CaRuO$_3$ and ferromagnetic SrRuO$_3$ films and was attributed to the zero band-curvature of the Fermi surfaces in these materials \cite{gausepohl_hall-effect_1996}. Although a sign change of $R_0$ is well known for compensated semicondcutors it is unusual for a metal where the carrier density is independent of temperature.

\section{IV. Discussion}
The resistivity, magnetoresistance, and Hall effect clearly show the characteristic features of a ferromagnetic metal, i.e., a kink in $\delta \rho/\delta T$ at $T_{\rm C}$, a temperature-dependent MR, and an anomalous Hall effect much larger than the ordinary Hall effect. However, both germanide films show a qualitatively different temperature dependence of  the AMR and ordinary Hall coefficient $R_0$ compared to the silicide film although both compounds have the same hexagonal crystal structure. In particular, for Mn$_5$Ge$_3$C$_{x}$ both coefficients, AMR and $R_0$, show a sign change in a similar tempüerature range. The difference in the temperature dependences presumably arises from the substantially different electronic band structure in the vicinity of $E_{\rm F}$ of the Mn silicide and Mn germanide which seems to be independent of C doping. This might originate from the different lattice constants which affects the hybridization of the orbitals. The volume of the crystallographic unit-cell increases continuously from Mn$_5$Si$_3$C$_{0.8}$ to Mn$_5$Ge$_3$C$_{0.8}$ to Mn$_5$Ge$_3$ \cite{surgers_magnetism_2009} in line with an increasing temperature dependence of $R_0$. The sensitivity of the magnetic moment and the spin polarization, i.e., spin-split band structure, to interatomic distances and strain in Mn$_5$Ge$_3$ has been reported earlier \cite{forsyth_spatial_1990,panguluri_spin_2005,dung_strain-induced_2013}. 

In the following we propose a scenario for the concomitant sign changes of the AMR ratio and Hall coefficient $R_0$ in Mn$_5$Ge$_3$C$_{x}$ films. In the two-current model for strong ferromagnets, the size of the AMR depends on the intraband scattering of conduction electrons by nonmagnetic impurities and on the scattering of conduction electrons into the unoccupied states of the $d_{\downarrow}$ band close to the Fermi level $E_{\rm F}$ \cite{campbell_spontaneous_1970,campbell_transport_1982}. The AMR is often positive while a negative AMR as observed in Fe$_4$N has been taken as evidence for minority-spin conduction \cite{tsunoda_negative_2009}. In this context, the two-current model has been extended to take into account (i) scattering into unoccupied $d$ states of both spin components, (ii) spin mixing of the $d$ bands by spin-orbit scattering, and (iii) spin-flip scattering arising from spin-dependent disorder and magnons \cite{kokado_anisotropic_2012}. The ratio $\rho_{s\downarrow}/\rho_{s\uparrow}$ of the resistivities of two bands of conducting \textit{s}, \textit{p}, and \textit{d} states arising from scattering by nonmagnetic impurities is treated as a variable together with the spin-resolved components of the $d$-band DOS at $E_{\rm F}$, $N^d_{\downarrow}$ and $N^d_{\uparrow}$. The AMR ratio arises from  slight changes of the $d$ orbitals by the spin-mixing term due to SOI. Interestingly, the sign of the AMR ratio does not depend on the absolute value of spin-flip scattering rate but on the ratios $\rho_{s\downarrow}/\rho_{s\uparrow}$, $N^d_{\downarrow}/N^d_{\uparrow}$, and the dominant $s-d$ scattering process \cite{kokado_anisotropic_2012}. 

The DOS of the spin-split band structure of Mn$_5$Ge$_3$ and Mn$_5$Ge$_3$C$_{0.8}$ has been obtained from first-principle calculations \cite{panguluri_spin_2005,stroppa_spin_2008,slipukhina_simulation_2009}. At the Fermi level, the total DOS $N(E_{\rm F})$ is dominated by $N^d(E_{\rm F})$ of the Mn$_{1}$ and Mn$_{2}$ $d$-states with a lower $N^d_{\uparrow}$ than $N^d_{\downarrow}$. The Ge and C $p$ bands do not contribute to the transport directly, but the Mn states in the majority spin band are strongly hybridized with the Ge $4p$ states. Similarly, the C $2p$ states hybridize with the Mn$_{2}$ states leading to a shift of the Mn$_{2}$ peaks in the DOS towards $E_{\rm F}$ and to an increased $N^d(E_{\rm F})$ in both spin channels while the Mn$_{1}$ states are left almost unaffected \cite{slipukhina_simulation_2009}. The calculations yield $N_{\downarrow}/N_{\uparrow} \approx N^d_{\downarrow}/N^d_{\uparrow} \approx$ 1.5 - 2 at $E_{\rm F}$ and an exchange splitting $E_{ex} \approx$\, 2.5 eV \cite{panguluri_spin_2005,stroppa_spin_2008,slipukhina_simulation_2009}. By using $1/\rho_{s\uparrow(\downarrow)} \approx e^2 N_{\uparrow(\downarrow)}(E_{\rm F}) \left\langle v_{\rm F \uparrow(\downarrow)} \right\rangle^2 \tau$ with the appropriate values given in Refs. \onlinecite{panguluri_spin_2005,slipukhina_simulation_2009} we obtain $\rho_{s\downarrow}/\rho_{s\uparrow} \approx$ 0.3 at low temperatures, i.e., a higher conductivity of the minority-spin channel, akin to Fe$_4$N \cite{tsunoda_negative_2009}. A similar $\rho_{s\downarrow}/\rho_{s\uparrow}$ value is derived from the spin polarization $P$ = -0.42 measured by Andreev reflection \cite{panguluri_spin_2005}. For $\rho_{s\downarrow}/\rho_{s\uparrow} \approx$ 0.3 we obtain in the extended two-current model \cite{kokado_anisotropic_2012} a negative AMR ratio = -0.3 \%  for a dominant $s-d$ scattering contribution $\rho_{s\rightarrow d\downarrow}/\rho_{s\uparrow}$ = 0.2. The maximum negative AMR ratio is usually of the order of $-\gamma = -\frac{3}{4}(\lambda_{SO}/E_{ex})^2 = - 1$\,\% as found experimentally, corresponding to $\lambda_{SO}$ = 0.3 eV in the present case, which is in fair agreement with the rough estimate mentioned above \cite{ye_berry_1999}.

The Hall constant in the two-current model is $R_0/\rho^2 = (R_{0 \uparrow}/\rho_{\uparrow}^2 + R_{0 \downarrow}/\rho_{\downarrow}^2)$ where $R_{0 \downarrow}$ and $R_{0 \uparrow}$ are the temperature independent ordinary Hall coefficients of the spin down und spin up band, respectively, and $\rho_{\downarrow}$ and $\rho_{\uparrow}$ their total resistivities. Due to the metallic character of the material and the linear $\rho(T)$ dependence, see Fig. \ref{fig1}, it is reasonable to assume temperature-independent carrier densities $n_{\downarrow}$ and $n_{\uparrow}$ and, hence, constant $R_{0 \downarrow}$ and $R_{0 \uparrow}$. 

From the AMR of Mn$_5$Ge$_3$C$x$ at low $T$ we know that the $\downarrow$ channel dominates the transport ($\rho_{\downarrow} \ll \rho_{\uparrow}$) and $R_0 < 0$ requires $R_{0 \downarrow} < 0$. At higher temperatures $R_0 > 0$ requires $R_{0 \uparrow}>0$ and $\rho_{\downarrow} > \rho_{\uparrow}$. Hence, $\rho_{\downarrow}/\rho_{\uparrow}$ must increase with $T$ in order to induce a sign change of the AMR \textit{and} $R_0$. The change from an electron-like minority-spin transport to a hole-like majority-spin transport in Mn$_5$Ge$_3$C$_{x}$ is possible since in a ferromagnetic metal electrons of one spin direction may constitute an electron-like Fermi surface while electrons of opposite sign may constitute a hole-like surface \cite{reed_high-field_1964}. 

From the extended two-current model an increase of $\rho_{\downarrow} / \rho_{\uparrow}$ with increasing temperature can be due to an increase of $\rho_{s\downarrow} / \rho_{s\uparrow}$ and/or of $N^d_{\uparrow} / N^d_{\downarrow}$ \cite{kokado_anisotropic_2012}. The latter has been proposed as an explanation for the AMR sign change in the half-metallic ferromagnet Fe$_3$O$_4$ with spin-split $t_{2g}$ and $e_g$ states \cite{kokado_anisotropic_2012,ziese_spontaneous_2000}. However, negative as well as positive ordinary Hall coefficients have been reported for Fe$_3$O$_4$ in the range $160 < T < 300$\, K \cite{reisinger_hall_2004,siratori_attempt_1988}. In half-metallic CrO$_2$ the electrons determine the conductivity while highly mobile holes determine the low-field magnetotransport properties \cite{watts_evidence_2000}. In the present case of conductive \textit{s}, \textit{p}, and \textit{d} states it is likely that the strong hybridization between the Mn 3\textit{d} states and the Ge states changes the conductivity of the spin-split conduction channels. This seems not to be the case for Mn$_5$Si$_3$C$_{0.8}$.  

\section{Summary}
In conclusion, we have investigated the Hall effect and anisotropic magnetoresistance of ferromagnetic Mn$_5$Ge$_3$, Mn$_5$Ge$_3$C$_{0.8}$, and Mn$_5$Si$_3$C$_{0.8}$ films. While for Mn$_5$Si$_3$C$_{0.8}$ the Hall coefficients are roughly independent of temperature, for Mn$_5$Ge$_3$C$_{x}$ these coefficients show a concomitant sign change from negative to positive in the same range of $T$. This is due to the fact that the electronic and magnetic properties in these Mn compounds depend very sensitively on the interatomic distances and hybridization. Hence, we have demonstrated a clear relation between the temperature dependence of the Hall coefficient and anisotropic magnetoresistance. Further work should show if this relation holds for other classes of ferromagnetic transition-metal compounds as well.  

We gratefully acknowledge financial support from the DFG Center for Functional Nanostructures (CFN). We thank I. A. Fischer for valuable discussions.

\end{document}